\documentclass[preprint,nofootinbib]{revtex4}
\usepackage{epsfig}
\usepackage{amssymb}

\begin{document}
\begin{titlepage}

\begin{center}
{\Large \bf A viable Starobinsky-like inflationary scenario in the light of Planck and BICEP2 results\footnote{Essay selected with honorable mention from Gravity Research Foundation (2014) - Awards Essays on
Gravitation}}


\vskip 0.5cm {\bf Spyros  Basilakos$^{a}$, Jos\'e  Ademir Sales
Lima$^{b}$, Joan Sol\`a$^{c}$}
\end{center}

{\small
\begin{quote}
\begin{center}
$^a$Academy of Athens, Research Center for Astronomy and Applied
Mathematics, Soranou Efesiou 4, 11527, Athens, Greece, e-mail: svasil@Academyofathens.gr \\
$^b$Departamento de Astronomia, Universidade de S\~ao Paulo, Rua
do Mat\~ao 1226, \\ 05508-900, S\~ao Paulo, SP, Brazil, e-mail: jas.lima@iag.usp.br\\
$^c$High Energy Physics Group, Departament d'Estructura i Constituents de la Mat\`eria, and Institut de
Ci\`encies del Cosmos, Universitat de Barcelona,\\ Av.
Diagonal 647 E-08028 Barcelona, Catalonia, Spain, e-mail: sola@ecm.ub.edu\\
\end{center}
\end{quote}
\vspace{0.2cm}
\centerline{(Submission date: March 31, 2014)}
}
\vspace{0.2cm}
\centerline{\bf Abstract}
\bigskip
The recent CMB data from Planck and BICEP2 observations have
opened a new window for inflationary cosmology. In this  Essay
we compare three  Starobinsky-like inflationary scenarios: (i) the
original Starobinsky proposal; (ii) a family of dynamically
broken SUGRA models; and (iii) a class of ``decaying'' vacuum
$\Lambda(H)$ cosmologies.  We then focus on the $\Lambda(H)$ variant, which
spans the complete cosmic history of the universe from an early
inflationary stage, followed by the ``graceful exit'' into the
standard radiation regime, the matter epoch and, finally, the
late-time accelerated expansion.  Computing the effective potential
we find that the  ``running'' $\Lambda(H)$ models also provide a
prediction for the tensor-to-scalar ratio of the CMB
spectrum, $r \simeq 0.16$, which is compatible
to within $1\sigma$ with the value  $r=0.20^{+0.07}_{-0.05}$
recently measured by the BICEP2 collaboration.

\vspace{0.3cm}

\end{titlepage}

\pagestyle{plain} \baselineskip 0.75cm

After decades of the successful big-bang paradigm, cosmology still lacks a framework in which the early inflationary phase of the universe smoothly matches the radiation epoch and evolves to the current quasi-de Sitter stage. Indeed,  in the light of the
Planck and BICEP2 results\,\cite{Planck,Bicep2}, a heated debate is taking place in the literature
about the best implementation of the inflationary paradigm\cite{ISL13,GKN14}.

Ijjas, Steinhardt and Loeb (ISL, \cite{ISL13})
claimed that there is a new
theoretical puzzle termed the {\it `unlikeliness problem'}. It comes from the fact
that only models with plateau potentials are favored
by Planck, but such scenarios are generically plagued  with
problems. For example,  the initial smoothness must be much larger than the Hubble volume, and this means that
inflation can only begin if the universe
is extremely  homogeneous. In addition,
there is a long debate on the
production of multi-universes and their measure\,\cite{multi}.

Such criticisms were  answered in a recent paper
by Guth, Kaiser and Nomura (GKN, \cite{GKN14}).  Despite they agreed that some criticisms of the ISL paper
are correct (for instance the Multiverse Problem\cite{multi}), their basic conclusion is that inflation
is a robust scenario which is on `strong footing than ever'.

The lively debate is going on, but in the meanwhile crucial new data also came to shed some light on it. This is precisely the context of our Essay. Here we focus on  some variants of Starobinsky's inflation\,\cite{staro} that turn out to be quite promising in the light of the data by Planck and the hot recent one from BICEP2\,\cite{Planck,Bicep2}.

\vspace{0.5cm}
{\it The original Starobinsky Inflation}:
It is the framework that implements de Sitter (inflationary) cosmological
solution of the gravitational field equations including higher curvature terms $\propto R^{2}$ in the effective action \cite{staro}:
%
%
\begin{eqnarray}\label{staroaction}
{\mathcal S} = \frac{1}{2 \, \kappa^2 } \, \int d^4 x \sqrt{-g}\,  \left(R  + \beta  \, R^2 \right) ~,~
\beta = \frac{8\, \pi}{3\, {\mathcal M}^2 }~,
\end{eqnarray}
where $R$ is the Ricci scalar, $\kappa^2=8\pi G$
and ${\rm G}=1/M_P^2$ is Newton's (gravitational) constant in
four space-time dimensions, with $M_P$ the Planck mass, and ${\mathcal M}$
a characteristic mass scale of the model.
The important feature of this model is that inflationary dynamics is
driven by the purely gravitational sector, through the $R^2$ terms,
and the scale of inflation is linked to ${\mathcal M}$.
However, the manifest disagreement of the BICEP2 measurement\,\cite{Bicep2} of the tensor-to-scalar ratio $r\simeq 0.2$ with the very small value  $r= 12/N^2$ ($N\simeq 50$ being the number of e-folds) predicted by the original Starobinsky model, has
triggered new formulations.

%
\vspace{0.5cm}
{\it Dynamically Broken SUGRA:}
The compatibility of the
dynamical breaking of supergravity (SUGRA) theories
via gravitino condensation
with Starobinsky~\cite{staro} inflationary scenario was also recently discussed\,\cite{ahmstaro}. Dynamical breaking of SUGRA, in the sense of the generation
of a mass for the gravitino field, whilst
the gravitons remain massless, occurs in the model as a
result of the four-gravitino interactions.
The one-loop effective potential, obtained by integrating
out gravitons and (massive) gravitino fields in the
scalar channel, may be
expressed as a power series in $\Lambda$.
The terms of order $\Lambda^2$ combine into
curvature scalar square terms, and the effective
action reads
%
%
\begin{equation}\label{effactionl3}
S_{SUGRA}\simeq \frac{1}{2\kappa^2} \int d^4 x \sqrt{-g}
\left(\alpha_1 \, \widehat R+ \alpha_2 \, \widehat R^2\right)~,
\end{equation}
with $\widehat R$ denoting the fixed $S^4$ background one expands around ($\widehat R=4\Lambda$, Volume = $24\pi^2/\Lambda^2$), and
the $\alpha$'s indicate the graviton
and gravitino quantum corrections at each order in $\Lambda$.
Obviously, one can see a link between the
action (\ref{effactionl3}) with a Starobinsky type action (\ref{staroaction}),
through an effective scale
$\beta_{\rm eff} \equiv  \alpha_2/\alpha_1$.

\vspace{0.5cm}
{\it Running Vacuum:}
It is remarkable that the
Cosmological Principle (embodied in the FLRW metric) does not prevent
$\Lambda$ to evolve with cosmic time or
a function of it. Thus, in the
Einstein equations,
$
R_{\mu \nu }-(1/2) g_{\mu \nu }R+\Lambda\,g_{\mu\nu}=-8\pi G\, {T}_{\mu\nu}\,,
$
%
%
the $\Lambda$-term can vary with time on equal footing to the matter density or the scale factor of the FLRW metric.
In fact the suggested time evolution of $\Lambda$ is indirect through the Hubble function $H(t)$. Specifically, the quantum field theory (QFT) in curved space-time, in combination with the renormalization group (RG), singles out
the possible general form of the evolution of the vacuum
energy density, $\rho_{\Lambda}=\Lambda/(8\pi G)$, as a function of $H(t)$ -- acting here as the natural running scale of the cosmic evolution. It suggests a RG-equation in which the rate of change of $\rho_{\Lambda}$ with
$H(t)$ contains only even powers of $H$ (because of the covariance of the
effective action)\,\cite{ShapSol2000}:
%
%
\begin{equation}\label{seriesRLH}
\frac{d\rho_{\Lambda}}{d\ln
H^2}=\frac{1}{(4\pi)^2}\sum_{i}\left[\,a_{i}M_{i}^{2}\,H^{2}
+\,b_{i}\,H^{4}+c_{i}\frac{H^{6}}{M_{i}^{2}}\,+...\right] \,,
\end{equation}
where the (dimensionless) coefficients receive loop contributions
from boson and fermion (hereafter $b$ and $f$) matter fields of
different masses $M_i$. There are no $M_i^4$ terms on the r.h.s. of (\ref{seriesRLH}) as $H<M_i$ for all particles and hence the RG prevents those terms from appearing\,\cite{ShapSol2000}. Obviously the expansion
converges very fast at low energies, where $H$ is rather small --
certainly much smaller than any particle mass. No other term beyond
$H^2$ (not even $H^4$) can contribute significantly on the
\textit{r.h.s.} of equation (\ref{seriesRLH}) at any stage of the
cosmological history below the GUT scale $M_{GUT}$, typically a few
orders of magnitude below the Planck scale $M_P\sim 10^{19}$ GeV. In contrast, in the very early universe (when $H$ is close, but below, the
masses of the heavy fields $M_i\sim M_{GUT}$), the $H^4$ effects can
be significant, in fact dominant. Integrating the above equation with only one leading high power term of $H$, generically called $H^{n+2}$ ($n\geq 1$), we arrive at
\begin{equation}\label{lambda}
\Lambda(H) = c_0 + 3\nu H^{2} + 3\alpha\,\frac{H^{n+2}}{H_{I}^{n}}\,.
\end{equation}
%
%
Here $c_0$ is an integration constant and $H_{I}$ is the Hubble parameter at the inflationary scale generated by the $H^{n+2}$ term.
The coefficients read: $\nu=\frac{1}{6\pi}\,
\sum_{i=f,b} c_i\frac{M_i^2}{M_P^2}$, $\alpha=\frac{1}{12\pi}\,
\frac{H_I^2}{M_P^2}\sum_{i=f,b} b_i$. They receive contributions from all the
matter particles and play the role of one-loop $\beta$-functions for the
RG running. Both coefficients are predicted to be naturally small since
$M_i^2\ll M_P^2$ for all the particles, even for the heavy fields of a
typical GUT below the Planck scale). In the case of $\nu$ an estimate
within a generic GUT is found in the range
$|\nu|=10^{-6}-10^{-3}$\,\cite{Fossil09},  compatible with observations\,\cite{BPS09}.

The equations of state for the dynamical vacuum and matter fluids are
still $p_\Lambda(t)=-\rho_\Lambda(t)$ and $p=\omega \rho$ ($\omega$
constant), respectively. The overall conservation law  in the
presence of a dynamical $\Lambda$-term reads
$\dot{\rho}+3(1+\omega)H\rho=-\dot{\rho_{\Lambda}}$, entailing energy
exchange between matter and vacuum. Combining this conservation
equation with (\ref{lambda}) and Friedmann's equation in flat space, $8\pi
G (\rho + \rho_{\Lambda}) = 3 H^2$, we obtain the differential evolution
law for $H(t)$. Let us express these results for  $n=2$ ($H^4$-inflation), which is the basic model. We find:
\begin{equation}
\label{HE}
\dot H+\frac{3}{2}(1+\omega)H^2\left[1-\nu-\frac{c_0}{3H^2}-
\alpha\left(\frac{H}{H_I}\right)^{2}\right]=0 \;.
\end{equation}
%
%
It admits the constant solution
$H=H_I[(1-\nu)/\alpha]^{1/2}$, corresponding to an inflationary
regime in the very early universe, i.e. when $H^2\gg c_0$.
The solution
of Eq.(\ref{HE}) is completely analytical, it suffices to say that
$
 H(a)/H_I\propto\,\left[1+D\,a^{4(1-\nu)}\right]^{-1/2}
$
in the radiation epoch ($\omega=1/3$),
where $D>0$ is a constant of integration.
For $Da^{4(1-\nu)} \ll 1$ the universe starts without
a singularity from the mentioned inflationary phase
powered by the huge value $H_I$, presumably connected to the scale of
a Grand Unified Theory (GUT) or even the Planck scale $M_P$.
At this epoch we have $\Lambda(H)\simeq 3\alpha H^{4}/H^{2}_{I}$ and
${\dot H}=0$, whereas the Ricci scalar becomes $R=12H^{2}$.
From the foregoing we may roughly expect that the pure gravitational part of the  underlying effective action should behave approximately as
\begin{equation}\label{steps1}
	S_{R,\Lambda} \simeq
\frac{1}{2\kappa^{2}}\int d^4 x \sqrt{-g}\,  \left(R+{\tilde \beta}R^{2} \right),
\;\;{\tilde \beta}=-\frac{\alpha}{24H^{2}_{I}}.
\end{equation}
%
%
Here  ${\tilde \beta}$ plays the role of an effective Starobinsky
coefficient: for ${\tilde \beta} \equiv \beta$
the time varying vacuum model (\ref{steps1}) can ``effectively'' be compared to
the Starobinsky inflationary model (\ref{staroaction}),
while for ${\tilde \beta}\equiv \beta_{\rm eff}$
it bares relation with
dynamically  broken SUGRA inflationary scenario (\ref{effactionl3})\, \cite{emdyno}. Of course there is no equivalence among these models, but they share some similarities.

The differences can actually be significant and may prove advantageous in some cases. In fact, our main question in this work is the following:
{\it what about the truly distinctive bonuses of the $\Lambda(H)$ class?} There are quite a few.

Firstly,
for $Da^{4(1-\nu)} \gg 1$ the universe evolves smoothly into essentially
the standard radiation phase $H(a) \simeq a^{-2(1-\nu)}$ ($|\nu|\ll 1$) and therefore successfully implements the ``graceful exit'' from the de Sitter stage. Then,
as the expansion proceeds, the  radiation component becomes sub-dominant, giving rise to the matter-dominated era. At this point the $H^{n+2}$-term in (\ref{lambda}) becomes negligible. Assuming flat $3$-dimensional space and integrating Eq.(\ref{HE})
we obtain $H^{2}(a)/{H_0^2}= \left[(1-\Omega_{\Lambda}^{0})\,a^{-3(1-\nu)}+\Omega_{\Lambda}^0-\nu \right]/({1-\nu})$.
This low-energy model turns out to be in agreement with the latest cosmological data,
including the growth rate of clustering -- basically as in
the $\Lambda$CDM\, \cite{runningH0,BPS09}. Not only so, a complete cosmological description emerges where the universe evolves between two extreme (early and late-time) de Sitter accelerating stages, with the ratio of the corresponding densities given by $\rho_f/\rho_I \sim (H_0/H_I)^{2} \sim 10^{-120}$ (for $H_I^{-1}$ close but below the Planck time). See e.g. the detailed discussion in \cite{runningH0}, as well as a plot of the complete cosmic history in Fig. 1 of \cite{HMention2013}.

Finally, we compute the tensor-to-scalar ratio ($r$) and the spectral
index ($n_s$) encoded in the CMB map within the effective potential approximation  associated to the model (\ref{lambda})\,\cite{HMention2013}. The potential can be expressed as follows:
\begin{equation}\label{Veffphi}
\frac{V_{\rm eff}(\phi)}{V(0)}=\frac{1+(1/3)\sinh^2\left(n\phi/2\right)}{\left[1+\sinh^2\left(n\phi/2\right)\right]^{(n+2)/n}}\,.
\end{equation}
The standard slow-roll parameters $(\epsilon,\eta)$ can now be readily calculated, and
from them $(n_s,r)$ immediately ensue. To this effect we perform a Taylor expansion of the effective potential around $\phi=0$, keeping terms up to $\phi^{2}$ and using the fact that $V''(0)/V(0)=-\,n\,(n+3)/3$.
The final result for the CMB parameters within this approximation reads:
\begin{equation}\label{eq:rANDns}
r=16\epsilon  \simeq 32N\,\left[\frac{n(n+3)}{f_{N}-3}\right]^{2}\,,\ \ \ n_{s}-1=-6\epsilon+2\eta \simeq -\frac{2n(n+3)(2f_{N}+3)}{(f_{N}-3)^{2}}\,,
\end{equation}
%
%
%
where
$f_{N}\equiv 2Nn(n+3)$, with $N$ the number of e-folds.
Using $N\simeq 50$ and $n=2$ (corresponding to $H^4$-inflation, which is the minimal model compatible with general covariance of the effective action) we obtain  $r\simeq 0.161$. This result is consistent
with the value $r=0.20^{+0.07}_{-0.05}$
that has been recently measured by the BICEP2 collaboration\,\cite{Bicep2}.

Within this approach we find  an universal prediction that is valid for virtually all $H^{n+2}$-models ($n\geq 1$), to wit: $r\to r^{u}\simeq 8/N\simeq 0.16$ 
and $n_s\to n_s^u\simeq 1-2/N\simeq 0.96$  (for $N=50$), a result 
which is in  agreement with the $1\sigma$ BICEP2 measurements\,\cite{Bicep2}.

We remark that $n_s^u$ is exactly as in the original 
Starobinsky's model. However, the predicted $r$ in that 
model is too small, $r=12/N^2\simeq 4.80\times 10^{-3}$ 
for $N=50$. Our value of $r^{u}$ is $2N/3\gtrsim33$ times 
bigger and can be in better agreement with observation.

To summarize, the inflationary class  $H^{n+2}$ of models (\ref{lambda}) seems to accomplish three main achievements $\forall n\geq 1$: i) it provides ``graceful exit''; ii) it smoothly connects the early universe with the $\Lambda$CDM model; and iii) it leads  to a prediction compatible with the  CMB parameters  $n_s$ and $r$  as measured by the combined Planck, WMAP, BAO  and BICEP2 data: $n_{s}=0.9607\pm 0.0063$ and $r=0.20^{+0.07}_{-0.05}$\,\cite{Planck,Bicep2}.


\end{document}